\documentclass[aps,prb,amssymb,showpacs,twocolumn]{revtex4-1}

\newcommand{\bA}{{\bf A}}

\newcommand{\bq}{{\bf q}}

\newcommand{\bp}{{\bf p}}
\newcommand{\bE}{{\bf E}}
\newcommand{\bv}{{\bf v}}

\newcommand{\br}{{\bf r}}
\newcommand{\brp}{{{\bf r}^{\prime}}}

\newcommand{\vf}{v_{\rm F}}

\newcommand{\sgn}{{\rm sgn}}

\newcommand{\la}{\lambda}

\newcommand{\rs}{\frac{e^2}{\varepsilon \lb}}
\newcommand{\sqpi}{\sqrt{\frac{\pi}{2}}}

\newcommand{\uz}{\hat{\bf u}_z}

\newcommand{\lb}{l_B}

\newcommand{\beq}{\begin{equation}}
\newcommand{\beqn}{\begin{eqnarray}}
\newcommand{\eeq}{\end{equation}}
\newcommand{\eeqn}{\end{eqnarray}}
\newcommand{\nn}{\nonumber}

\usepackage{graphics,graphicx,amsmath}

\usepackage{tabularx,float}
\usepackage{epsfig}

\usepackage{subfigure}

\usepackage{bm}
\usepackage{wasysym}

\usepackage{bm}
\usepackage{color}
\usepackage{verbatim}

\begin{document}

\bibliographystyle{apsrev4-1}

\date{\today}

\author{R. Rold\'{a}n$^{1,2}$, J.-N. Fuchs$^2$ and M. O. Goerbig$^2$}
\affiliation{\centerline{$^1$Institute for Molecules and Materials, Radboud University Nijmegen, Heyendaalseweg 135, 6525 AJ Nijmegen, The Netherlands}\\
\centerline{$^2$Laboratoire de Physique des Solides, Univ. Paris-Sud, CNRS, UMR 8502, F-91405 Orsay Cedex, France}}

\title{Spin-flip excitations, spin waves, and magneto-excitons in graphene Landau levels at integer filling factors}

\begin{abstract}
We study collective electronic excitations in graphene in the
integer quantum Hall regime, concentrating mainly on excitations
with spin reversal such as spin-flip and spin-wave excitations. We
show that these excitations are correctly accounted for in the
time-dependent Hartree-Fock and strong magnetic field approximations, in contrast to
spin-conserving (magneto-exciton) modes which involve a strong
Landau-level mixing at non-zero wave vectors. The collective
excitations are discussed in view of prominent theorems, such as
Kohn's and Larmor's. Whereas the latter remains valid in graphene
and yields insight into the understanding of spin-dependent modes,
Kohn's theorem does not apply to relativistic electrons in graphene.
We finally calculate the exchange correction to the chemical
potential in the weak magnetic field limit.
\end{abstract}


\maketitle

\section{Introduction}

The role of electron-electron interactions in graphene
(two-dimensional graphite) is still a debated issue. Whereas most of
its electronic properties can be understood within a model of
two-dimensional (2D) noninteracting massless Dirac
fermions,\cite{CG07} there are some experimental indications
for the presence of Coulomb interactions.\cite{LB08} These
correlations, which may be quantified by the graphene fine-structure
constant $\alpha_G=e^2/\hbar \varepsilon \vf\simeq 2.2/\varepsilon$,
in terms of the Fermi velocity $\vf$ and the dielectric constant
$\varepsilon$, seem, however, to be weak and long-ranged. Therefore,
strongly-correlated phases that are expected in the large-$\alpha_G$
limit\cite{DL09} or for short-range Hubbard
interactions\cite{tosatti,herbut,lee} are unlikely to occur in
undoped or moderately doped graphene. Theoretically, a perturbative
Fermi-liquid-type treatment of the Coulomb interactions yields a
logarithmic divergence of the Fermi velocity,\cite{GGV94} a
renormalization of thermodynamic quantities such as the
compressibility\cite{BM07} as well as to a control of the orbital
magnetic susceptibility.\cite{PPVK10}

The situation is different if the graphene electrons are exposed to a strong magnetic field that quantizes their kinetic energy into
non-equidistant Landau levels (LLs), $\epsilon_n=(\lambda\hbar\vf/l_B)\sqrt{2n}$, where $\lambda=\pm 1$ is the band index, $l_B=\sqrt{\hbar c/eB}$ is
the magnetic length, and $n$ denotes the LL index.\cite{CG07} The most prominent consequence of this relativistic LL quantization
and the presence of a zero-energy LL for $n=0$ is a peculiar integer quantum Hall effect (QHE), with an unusual sequence
of Hall plateaus.\cite{NF05,ZTSK05} The LLs are highly degenerate, as in the usual 2D electron gas (2DEG) with a parabolic band
dispersion, where the density of states per LL (and per unit area) is given by the flux density $n_B=1/2\pi l_B^2=eB/h$, which is
proportional to the perpendicular magnetic field. The filling of the LLs is then characterized by the ratio (filling factor)
$\nu=n_{el}/n_B$ between the 2D electronic density $n_{el}$ and $n_B$.
A partially filled LL may then be viewed as a strongly-correlated electron
system with a quenched kinetic energy, and its most prominent manifestation is a fractional QHE that has recently been
observed in suspended graphene samples.\cite{DA09,BK09} Prior indications for strong interactions stemmed from
the high-field QHE at $\nu=0$, which indicates a stronger lifting of the four-fold spin-valley degeneracy of graphene than
what one would expect from single-particle effects.\cite{ZK06,GZ09}

Similarly to the $B=0$ case, the Coulomb interaction between electrons in completely filled LLs may be viewed as a weak perturbation
because of the energy gap between adjacent LLs. Its role in the dispersion relation and Fermi velocity renormalization of graphene is an open question which has been addressed both theoretically\cite{IWFB07,BM08,S07} and experimentally, in
the framework of transmission spectroscopy.\cite{SH06,JS07,DG07,HS10,OP10} As compared
to the 2DEG with a parabolic band dispersion (as in GaAs/AlGaAs and Si/SiGe heterostructures), the situation is strikingly different in graphene, where
the effect of electron-electron interactions may be probed at zero wave vector. Indeed, in the former,
LL quantization leads to a set of equidistant LLs separated by the cyclotron frequency.
Kohn's theorem states that in these systems, homogeneous electromagnetic radiation can only couple to the center-of-mass coordinate.
Therefore internal degrees of freedom associated with the Coulomb interaction cannot be excited by such optical probes.\cite{K61}
Then, the dispersion relation of spin-conserving magnetoplasmons at zero wave-vector is equal to the bare cyclotron energy,
irrespective of existing electronic correlations.\cite{KH84} A similar consideration holds also for spin wave (SW) modes,
for which Larmor's theorem states that the Coulomb interaction does not renormalize the zero-wave-vector dispersion of the
spin excitons.\cite{DKW88} However, the dispersion of spin-flip (SF) modes in a 2DEG are shifted from the bare cyclotron resonance
even at zero wave-vector, due to electron-electron interactions. Therefore, these excitations are the only suitable modes
to study the many-body effects in a 2DEG by means of optical measurements.\cite{PW92,KW01,VW06}

Furthermore, electron-electron interaction in the regime of the
integer QHE yield collective excitations that are different from
those in the 2DEG -- instead of inter-LL excitations with a rather
weak wave-vector dispersion, called magneto-excitons (ME), one finds
linear magneto-plasmons that involve superpositions of different LL
transitions.\cite{RFG09,RGF10} Also MEs that may play a role in the
vicinity of $q=0$ have been studied theoretically in graphene, and
it has been shown that Coulomb interactions yield a renormalization
of the transition energy at zero wave vector\cite{IWFB07,BM08,S10}
that indicate that Kohn's theorem does not apply to graphene.
In comparison to these works, here we put more emphasis on
spin-changing modes and on the effect of LL mixing.

A convenient way of assessing the magnetic-field strength is
in terms of four characteristic length scales: the magnetic length
$l_B$, the carbon-carbon distance $a$, the Fermi wavelength
$\lambda_F$ and the Thomas-Fermi screening length $\lambda_{TF}\sim
\lambda_F/\alpha_G$ where $\alpha_G\equiv e^2/\varepsilon v_F$
measures the relative strength of Coulomb interactions. In practise,
the magnetic length is always much larger than the lattice spacing
$l_B\gg a$ because the flux per unit cell is much smaller than the
flux quantum ($B\ll 40000$~T). The weak LL mixing approximation --
which we will use when studying the particle-hole excitations --
requires $e^2/(\varepsilon l_B \omega_C)\ll 1$, where $\omega_C$ is
the cyclotron frequency, and corresponds to a strong field such that
$l_B\ll \lambda_{TF}$. As in graphene $\alpha_G$ is of order 1,
$\lambda_{TF}\sim \lambda_F$ and the weak LL mixing is also the
small filling factor limit $l_B\ll \lambda_F$. In the following, we
will consider two limits: either a strong magnetic field (meaning
$l_B\ll \lambda_F$ or typically $B\gg 20$~T) or a weak magnetic
field (meaning $l_B\gg \lambda_F$ or typically $B\ll 20$~T).

In this paper, we study both spin-conserving ME and spin-dependent
SF and SW modes in the regime of the integer
QHE. Following the scheme introduced by Kallin and Halperin
(KH) for the 2DEG,\cite{KH84} the Coulomb interaction is treated
within the framework of the time-dependent Hartree-Fock (TDHF) 
and strong-field approximation [for which $e^2/(\varepsilon
l_B\omega_C)\ll 1$, which insures that LL mixing is weak],
the validity of which is discussed. Indeed, we find that whereas SF
and SW may be accounted for correctly in graphene within the KH approximation in the limit of a strong magnetic field and
when the Fermi level lies near the $n=0$ LL, its validity in the
treatment of spin-conserving ME is questionable even in these
limits. This difference between MEs and SF (and SW) excitations
stems from the depolarization term, accounted for in the
random-phase-approximation (RPA), which is present only in the ME
dispersion and which yields a strong LL mixing at non-zero values of
the wave vector. This LL mixing eventually leads to the formation of
linearly dispersing plasmon-type modes that have been obtained
within an RPA treatment of the electronic polarizability in
graphene.\cite{RFG09,RGF10}

Finally, we consider the opposite limit of graphene in a
weak magnetic field, and compute the exchange correction to the
chemical potential. We find that the exchange correction to the
single-particle dispersion presents the same dependence in the two
limits, strong and weak magnetic field, proportional to the square
root of the ultraviolet cutoff in the Landau levels. On the other
hand, the exchange correction to the particle-hole dispersion
diverges logarithmically with the cutoff.

The paper is organized as follows. In Sec. \ref{Sec:ExcMod}, we
first revisit the Kohn's (Sec. \ref{sec:Kohn}) and Larmor's (Sec.
\ref{sec:Larmor}) theorems in the context of graphene and their
impact on collective excitations in general. We then study the
excitonic modes in graphene in a strong magnetic field, within the
KH approximation. In Sec. \ref{Sec:MuEx} we calculate the
exchange correction to the chemical potential, in the weak magnetic
field limit and/or for highly doped samples. Our main conclusions
are summarized in Sec. \ref{Sec:Conc}, and the technical details of
the calculations are provided in the appendices.

\section{Excitonic modes in graphene in the integer QHE}\label{Sec:ExcMod}

\begin{figure}[t]
  \centering
   \includegraphics[width=0.3\textwidth]{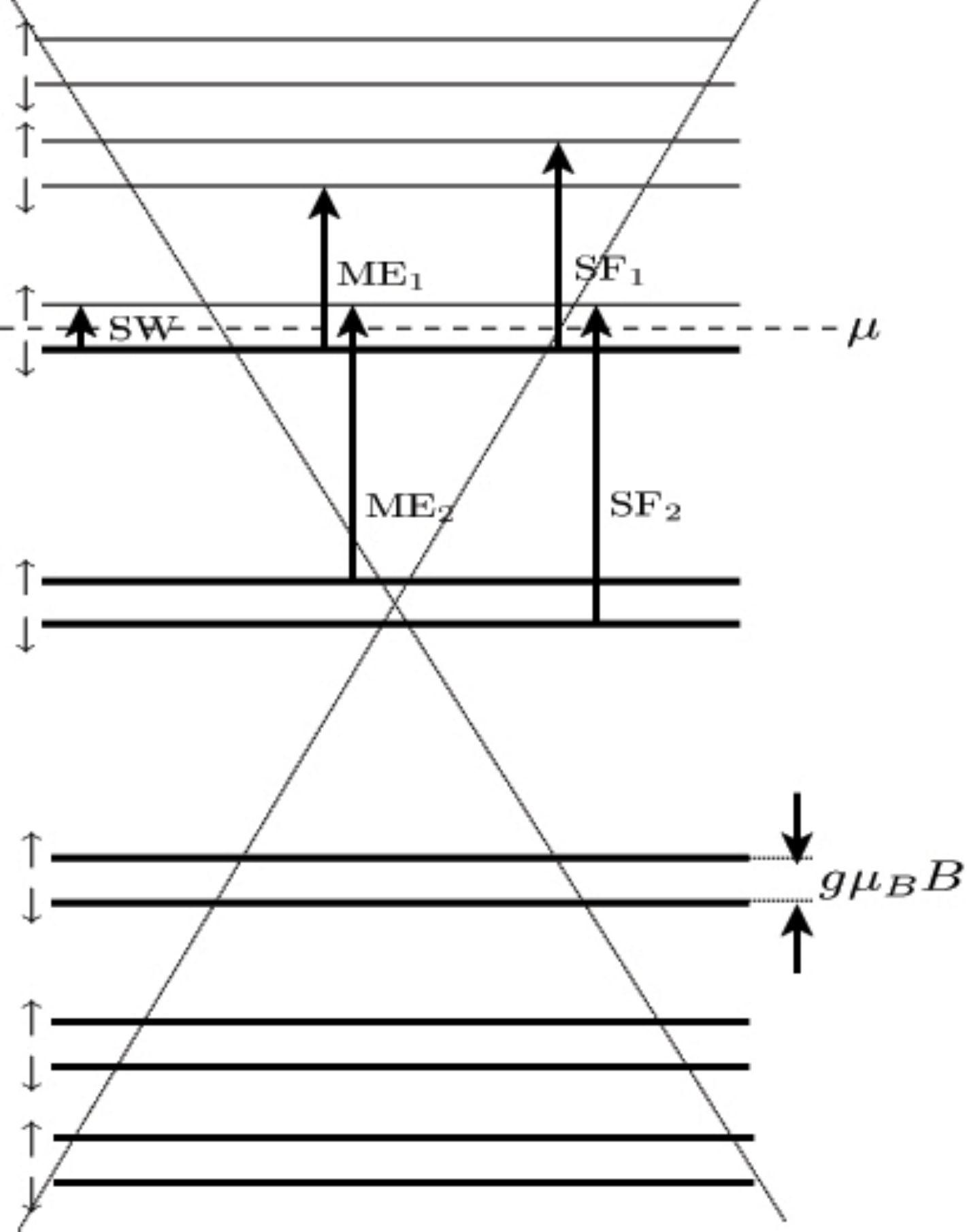}
\caption{Sketch of the particle-hole excitations studied in the text. Each Landau level is split in two sub-levels, separated by the Zeeman gap, $\Delta E_z=g\mu_BB$. Here, $\uparrow$ indicates $s_z=+1/2$ and $\downarrow$, $s_z=-1/2$. Label ME stands for magneto-exciton (or magneto-plasmon) where the particle and the hole have the same spin.  SF denotes the spin-flip excitation, in which the electron and the hole do not only reside in different LLs but also have different spin. Finally, SW denotes the spin-wave mode, which is an intra-LL transition where the electron and the hole have a different spin orientation and where we have taken into account a non-zero Zeeman energy.}
  \label{Fig:Trans}
\end{figure}

In spin-flip modes, an electron is both promoted from one LL to the next one and its spin is reversed. They carry a spin $S_z=s^e_z  - s^h_z=\pm1$, where $s^{e(h)}_z$ is the $z$-component of the electron (hole) spin. We use the term magneto-exciton (ME, sometimes also called magneto-plasmon in the literature) to denote spin-conserving excitations, where the electron and the hole reside in different LLs and have the same spin. In SW modes, the two particles have the same LL index but opposite spin. SF modes, MEs and SWs are the basic excitations of a quantum Hall system with a finite Zeeman splitting, as sketched in Fig. \ref{Fig:Trans}.

\subsection{The fate of Kohn's theorem in graphene}
\label{sec:Kohn}

Before calculating the dispersion relations of the different excitonic modes, we discuss here qualitatively the expectations for
graphene with respect to the 2DEG with a parabolic band dispersion. In the latter, Kohn's theorem\cite{K61} states that electromagnetic
absorption, irrespective of the strength of the Coulomb interaction, occurs only at
the cyclotron frequency $\omega_C=eB/m_b$, in terms of the band mass $m_b$.
Here and in the remainder of this paper, we consider a system of units with $\hbar\equiv c \equiv 1$. Because this absorption process
is associated with an inter-LL transition from the last occupied to the first unoccupied level, it means that the lowest-energy
ME must converge to the non-interacting value at zero wave-vector. This statement remains valid also for higher harmonics, such that
the ME dispersion has $\Omega_{ME}(q\rightarrow 0)\rightarrow m\omega_C$, where $m=n_e-n_h$ is the difference between the LL index of the electron ($n_e$) and that of the hole ($n_h$).

\subsubsection{Kohn's theorem in the 2DEG}

In order to investigate the fate of Kohn's theorem in graphene, let
us recall the main steps in the argument for the 2DEG. In the
latter, the Hamiltonian of $N$ non-interacting electrons can be
expressed in terms of  the gauge-invariant momenta
$\mathbf{\Pi}_j=\bp_j + e\bA(\br_j)$, where $\br_j$ and $\bp_j$ are
the position and its conjugate (gauge-dependent) momentum,
respectively, of the $j$-th electron with charge $-e$ (we choose
$e>0$ to be the {\sl positive} elementary charge), 
\beq\label{zeroHam2DEG} H_0=\frac{1}{2m_b}\sum_{j=1}^N
\mathbf{\Pi}_j^2. \eeq From the total gauge-invariant momentum
$\mathbf{\Pi}=\sum_j \mathbf{\Pi}_j$, we define the raising and
lowering operators $\Pi^{\pm}=\Pi_x \pm i \Pi_y$, which satisfy the
commutation relations $[\Pi^{\pm},H_0]=\mp (1/m_bl_B^2)\Pi^{\pm}$
with the Hamiltonian $H_0$. This leads to the equation
\beq\label{eigen}
H_0\left(\Pi^{\pm}\left|\psi^0\right\rangle\right)=\left(E^0 \pm
\omega_C\right)\left(\Pi^{\pm}\left|\psi^0\right\rangle\right), \eeq
which means that the application of $\Pi^{\pm}$ on an ($N$-particle)
eigenstate $|\psi^0\rangle$ (with energy $E^0$) of $H_0$ yields
another eigenstate with energy $E^0 \pm \omega_C$.

The first observation is that this equation remains valid also in
the presence of electron-electron interactions $V$ that
commute with the total momentum $[\mathbf{\Pi},V]$=0, such
as the Coulomb interaction, if one replaces the non-interacting
state $|\psi^0\rangle$ by an eigenstate $|\psi\rangle$ of the full
Hamiltonian $H=H_0+V$, as well as the energy $E^0$ by that,
$E$, of the state $|\psi\rangle$.

Second, one notices that the electromagnetic light field with
frequency $\omega$ couples to the electronic system via the
Hamiltonian 
\beq\label{LMcoupl} H_{LM}(t) =
\frac{e}{2i\omega} e^{-i\omega t} \bE(\omega)\cdot \sum_j \bv_j +
\textrm{ H.c.} , \eeq 
where $\bE(\omega)$ is the electric component
of the light field and $\bv_j$ the velocity operator of the $j$-th
electron. In the 2DEG with a parabolic band dispersion, the velocity
operator is readily expressed in terms of the gauge-invariant total
momentum, $\sum_j \bv_j=\mathbf{\Pi}/m_b$, such that the
light-matter coupling (\ref{LMcoupl}) is linear in the operators $\Pi^{\pm}$.
As mentioned above, this induces then a transition from a state $|\psi\rangle$ with energy $E$ to a state $\Pi^{\pm}|\psi\rangle$
with energy $E\pm \omega_C$, i.e. the only absorption peak for light occurs at the cyclotron frequency $\omega_C$.\cite{K61}

\subsubsection{Difference in graphene}
Although also in graphene the total gauge-invariant momentum
$\mathbf{\Pi}$ commutes with the interaction Hamiltonian $V$
but not with $H_0$, one first notices that it may no longer be
expressed in terms of the velocity operators of (now
relativistic) electrons because of the vanishing band mass. The
velocity operator is a $2\times 2$ matrix $\bv_j=\vf
\boldsymbol{\sigma}_j=\vf (\sigma_j^x,\sigma_j^y)$, in terms of the
Pauli matrices $\sigma^x$ and $\sigma^y$, and it is not a conserved
quantity even in the absence of interactions. The application of the
velocity operator on an eigenstate of the non-interacting
Hamiltonian $H_0=\sum_j\vf
[\bp_j+e\bA(\br_j)]\cdot\boldsymbol{\sigma}_j$, for which
$[\bv_j,H_0]\neq 0$, yields, even in the absence of a magnetic field, spontaneous inter-band transitions that
are at the origin of the so-called {\sl zitterbewegung}.\cite{K06}
As a consequence, the light-matter coupling Hamiltonian
(\ref{LMcoupl}), the form of which is also valid for graphene, may
no longer be expressed in terms of $\Pi^{\pm}$. Indeed, the velocity
operator in Hamiltonian (\ref{LMcoupl}) yields transitions involving
LLs with adjacent indices $n$ and $n\pm 1$, as in the 2DEG, but the
{\sl zitterbewegung} translated to the magnetic-field
case\cite{RZ08} furthermore yields inter-band excitations, such that
the dipolar selection rules $\lambda_h,n\rightarrow \lambda_e,n\pm
1$ are associated with the energies
\begin{equation}\label{CyclRes}
E_{kin}^{(n,\lambda_e,\lambda_h)}=\frac{\vf}{l_B}\sqrt{2}\left(\lambda_e\sqrt{n+1}-\lambda_h\sqrt{n}\right),
\end{equation}
where one expects absorption peaks. Therefore, already in the non-interacting limit, one expects a plethora of absorption
peaks, that have indeed been observed experimentally,\cite{SH06,DG07,JS07,HS10} and not a single cyclotron resonance as in the
case of the 2DEG with a parabolic dispersion relation.

Furthermore, because the kinetic Hamiltonian (\ref{zeroHam2DEG}) becomes $H_0=\vf \sum_j\mathbf{\Pi}_j\cdot\boldsymbol{\sigma}_j$ in
graphene, one loses the possibility of writing an equation of the type (\ref{eigen}) for graphene, neither in
terms of the total momentum $\mathbf{\Pi}$ nor with the help of $\sum_j\bv_j$, which as we mentioned is not conserved.
There is thus no protection of the energies
(\ref{CyclRes}) when interactions are taken into account. Indeed, the latter renormalize the absorption energies, \cite{IWFB07,BM08,S10}
as we discuss below, in contrast to the 2DEG, where the absorption energy is protected by Kohn's theorem, and the
ME modes no longer converge to the non-interacting inter-LL transition energies
\begin{equation}\label{TransEn}
E_{kin}^{(n_e,n_h)}=\frac{\vf}{l_B}\sqrt{2}\left(\lambda_e\sqrt{n_e}-\lambda_h\sqrt{n_h}\right)
\end{equation}
in the zero-wave-vector limit.

\subsection{Larmor's theorem applied to graphene}
\label{sec:Larmor}

In addition to ME excitations that do not involve the electronic
spin, one may investigate spin excitations on rather general
grounds. Larmor's theorem states that in the long-wavelength limit,
the SW dispersion tends to the (bare) Zeeman splitting,
$\Omega_{SW}(q\rightarrow 0)\rightarrow g\mu_BB$.\cite{DKW88} This
theorem may be understood from the symmetries of the Hamiltonian
$H=H_0 + H_{int} + H_Z$. In the absence of the Zeeman term $H_Z$,
the Hamiltonian respects the SU(2) symmetry associated with the
electronic spin, i.e. both the total spin operator $\hat{S}_{tot}^2$
and any of the components $\hat{S}_{tot}^{\mu}$, for $\mu=x,y,z$,
commutes with the Hamiltonian. Since one cannot diagonalize all
components of the total spin simultaneously, one needs to choose a
particular one, and this is naturally the one chosen by the Zeeman
effect (here $\hat{S}_{tot}^z$), such that the full Hamiltonian
commutes with $\hat{S}_{tot}^2$ and $\hat{S}_{tot}^z$. The quantum
numbers associated with the spin, $S$ and $S^z$, are therefore good
quantum numbers for the full interacting $N$-particle Hamiltonian,
such that all possible states have energies $E=E(S,S^z, ...)
+ g\mu_B B S^z$, where the dots $...$ represent other quantum
numbers that characterize the interacting system. The essence of this expression is that the full interacting
$N$-particle system may be viewed as a large spin that precesses in 
a magnetic field with the fundamental (Larmor) frequency $\omega_L=g\mu_B B$.
Whereas this frequency is affected by the (crystalline) environment via
the effective $g$-factor, the latter remains unaltered by the 
electron-electron interactions.
Applied to the present problem of collective excitations, this means
that the Zeeman term does not represent a further complication to
the SU(2) symmetric Hamiltonian $H_0+ H_{int}$, which thus needs to
be diagonalized first.

These rather obvious considerations allow us to understand easily
Larmor's theorem if one notices that, in the absence of a Zeeman
effect, the SW mode is just the Goldstone mode of a ferromagnetic
ground state in which all spins are spontaneously polarized. This
ferromagnetic state arises due to exchange-interaction effects when
not all subbranches of a particular LL are completely
filled.\cite{MS95} The Goldstone mode is characterized by a
dispersion relation that vanishes (as $q^2$ for a SW mode
\cite{HH69}) in the zero-wave-vector limit, $\omega_G(q\rightarrow
0)\rightarrow 0$, which means that the different states of the
ground-state manifold (i.e. the different polarizations) are
connected by a global rotation of zero energy cost that is precisely
the $q=0$ Goldstone mode. In the presence of the Zeeman effect,
which chooses a particular orientation of the total spin, one thus
obtains a SW mode that tends to the energy $\Omega_{SW}(q\rightarrow
0)\rightarrow g\mu_B B S_z$, where $S_z=1$, as stated by Larmor's
theorem.

One notices that, in contrast to the above discussion of Kohn's theorem, the (non-)relativistic character of $H_0$ has never
played a role in the argument, and Larmor's theorem therefore also applies in the case of graphene. Moreover, one is confronted
in graphene with an additional two-fold valley degeneracy, that may be taken into account by an SU(2) valley isospin. Although
the SU(2) valley symmetry is not respected by the interaction Hamiltonian, the symmetry-breaking terms are strongly suppressed
(by a factor of $a/l_B\sim 0.005\sqrt{B\text{[T]}}$, in terms of the carbon-carbon distance $a=0.14$ nm) such that the interaction
Hamiltonian is approximately SU(2) valley-symmetric.\cite{GMD06,G10} Therefore the above arguments apply also to possible valley-ferromagnetic
states in graphene, i.e. there are valley-isospin-wave modes that vanish in the $q\rightarrow 0$ limit and that may become
eventually gapped by a ``valley-Zeeman'' effect $H_{v-Z}$ that, if it may be written in terms of components of the total valley-isospin,
yields a simple energy offset to the dispersion. In the remainder of the paper, we concentrate on collective excitations that involve
only the physical spin.

In addition to this generalization of Larmor's theorem to the valley isospin, it may also be generalized to the
SF modes, which involve not only different spin states but also different LLs. The dispersion of the collective SF modes may be fully understood from the
Hamiltonian $H_0 + H_{int}$, whereas the energy $g\mu_BB S_z$ associated with the Zeeman effect can be simply added at the end
of the calculation as a global (wave-vector independent) constant. However, as we shall discuss below, the energy of the
SF modes does not converge to the simple sum of the Zeeman and the transition energies (\ref{TransEn}) in the zero-wave-vector
limit, but they are renormalized by the interaction energy, both in graphene and in the 2DEG.\cite{PW92}

\subsection{Dispersion relation of the excitonic modes}\label{Sec:KHappr}

In graphene, the energies of ME, SW and SF modes can be expressed as:
\begin{eqnarray}
\Omega_{ME}(q)&=&E_{kin}^{(n_e,n_h)}+\Delta E^{(n_e,s_z^e;n_h,s_z^h)}(q)\label{Eq:ME}\\
\Omega_{SW}(q)&=&g\mu_BBS_z+\Delta E^{(n_e,s_z^e;n_h,s_z^h)}(q)\label{Eq:SW}\\
\nn
\Omega_{SF}(q)&=&E_{kin}^{(n_e,n_h)}+g\mu_BBS_z+\Delta E^{(n_e,s_z^e;n_h,s_z^h)}(q)\\
\label{Eq:SF}
\end{eqnarray}
where $S_z$ is the $z$-component of the exciton spin, and $E_{kin}^{(n_e,n_h)}$ is the
transition energy in the absence of interactions given by Eq.
(\ref{TransEn}). The contribution $\Delta E^{(n_e,s_z^e;n_h,s_z^h)}$
consists of three terms (see Appendix \ref{App:Poles} for details):
a depolarization or exchange term $E_{x}(q)$, which is accounted for
in the RPA approximation, a direct Coulomb interaction between the
electron and hole (vertex corrections) $E_v(q)$, and the difference
between the exchange self-energy of the electron and that of the
hole, $E_{exch}=\Sigma_e-\Sigma_h$. Notice that $E_{x}(q)$ is only
relevant for the ME, because only particles with the same spin can
be recombined by means of electron-electron interactions.

It must be kept in mind that, in a 2DEG, the RPA term, which
determines the maximum of the ME dispersion at a wave-vector $q\sim
1/R_C$ in the TDHF approximation, mixes different LLs, with a mixing
amplitude on the order of $e^2/(\varepsilon
l_B\omega_C)$.\cite{GV05}  Here $R_C=k_Fl_B^2$ is
the cyclotron radius, where the Fermi momentum in terms of the index $N_F$ of the topmost
fully occupied LL is $k_F=\sqrt{2N_F+1}/l_B$ for a 2DEG and $k_F=\sqrt{2N_F+\delta_{N_F,0}}/l_B$ for graphene. This needs to be distinguished from the LL mixing
at $q=0$, which determines the stability of the LLs in the presence
of electron-electron interactions and which scales as
$e^2/\varepsilon R_C \omega_C$. Although the stability of the LLs in
graphene is determined by the ratio between the Coulomb energy
$e^2/\varepsilon R_C$ and the LL separation $\Delta_n=
(\sqrt{2}\vf/l_B)(\sqrt{N_F+1}-\sqrt{N_F})\sim \vf/R_C$,\cite{G10}
which happens to be the scale-invariant fine-structure constant
$\alpha_G=e^2/\varepsilon \vf$, the situation is again different at
the maximum of the ME dispersion at $q\sim 1/R_C$. The order of
magnitude of the $q\neq 0$ LL mixing in graphene may be obtained by replacing in
$e^2/(\varepsilon l_B\omega_C)$ the 2DEG cyclotron frequency
$\omega_C=eB/m_b$ by the density-dependent cyclotron frequency
$\omega_C(\mu)=eBv_F^2/\mu$, where $\mu=(\vf/l_B)\sqrt{2N_F}$ is the
chemical potential. As a consequence, the validity of the KH
approximation fails not only for weak magnetic fields, as in the
standard 2DEG, but also at high and intermediate filling factors
because the effective cyclotron frequency in graphene decreases as
the number of filled LLs increases, leading to an increase of the LL
mixing. Therefore, strictly speaking, the results of this section
will be valid only in the strong-$B$ limit and for $N_F$  near 0.
However, we will see that the KH approximation can still be
applied for spin-dependent excitations (SW and SF) slightly away
from half-filling, but not to spin-conserving modes (ME). This is a
consequence of the absence of the depolarization term $E_x(q)$,
which is the main source of LL mixing, in SW and SF modes, whereas
it constitutes the main contribution to the dispersion of ME modes.

After these general considerations on collective excitations, we now turn to a discussion of the modes at particular integer
filling factors, which are described by
\beq\label{filling}
\nu=4N_F-2+2\left(\nu_{\uparrow}^{N_F} + \nu_{\downarrow}^{N_F}\right),
\eeq
where $N_F$ is the index of the top most fully occupied LL, $0\leq \nu_{\sigma}^n\leq 1$
is the filling of the spin-$\sigma$ branch of the $n$-th LL, and the factor of 2 accounts for the two-fold
valley degeneracy of each spin branch.

\begin{figure}[t]
  \centering
    \subfigure[]{\label{DisperNF0}\includegraphics[width=0.35\textwidth]{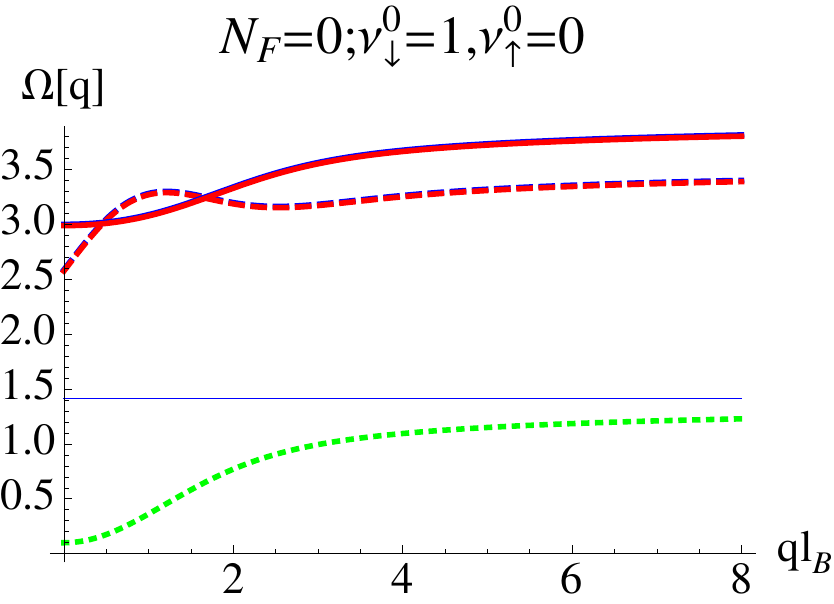}}
     \subfigure[]{\label{TransNF0}\includegraphics[width=0.35\textwidth]{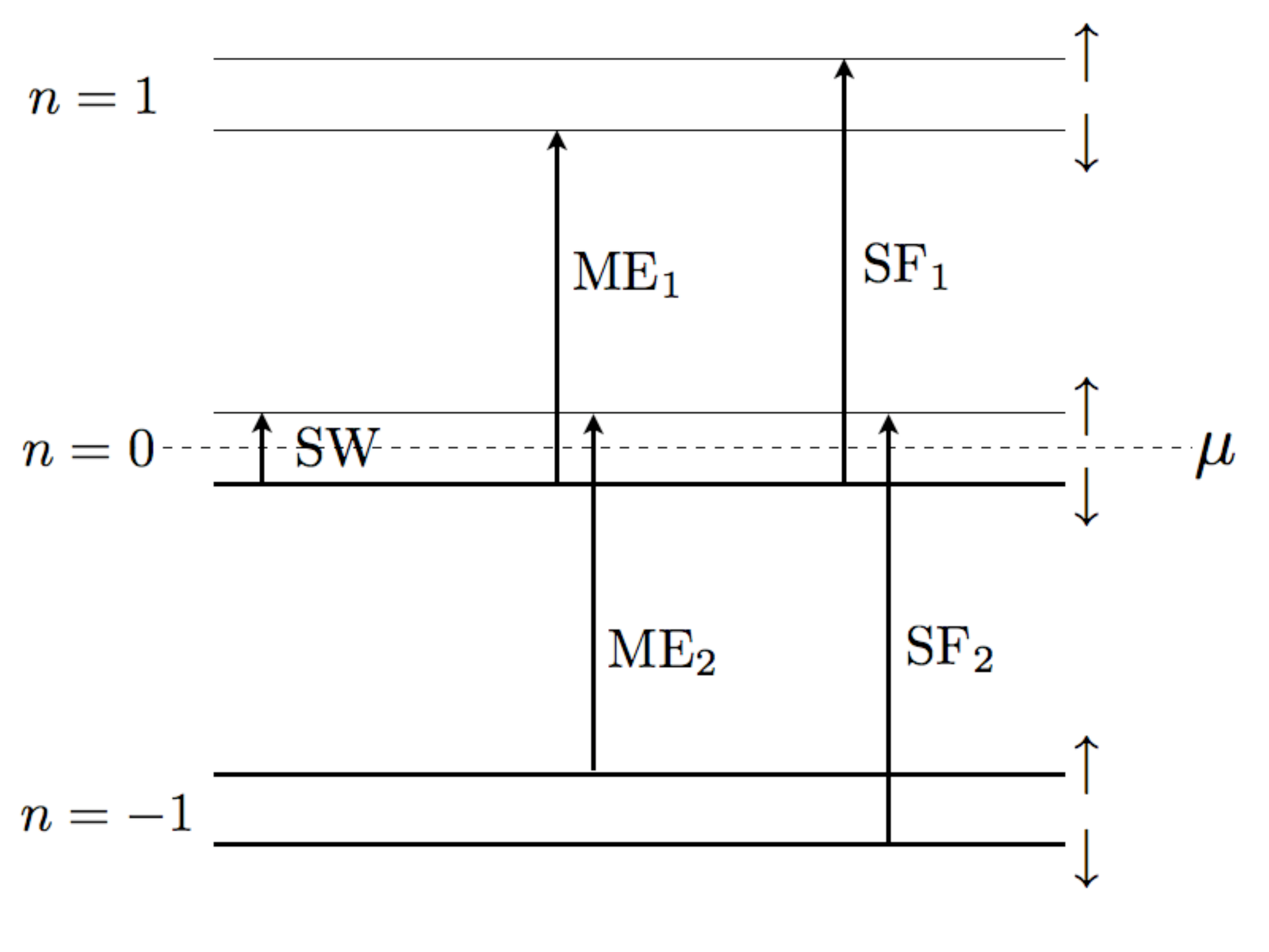}}
 \caption{Dispersions (in units of $e^2/\varepsilon \lb$) of the excitonic modes studied for $\nu=0$, i.e.
$N_F=0$, $\nu_{\downarrow}^0=1$ and $\nu_{\uparrow}^0=0$. SW (dotted
green line), ME$_{1,2}$ (dashed blue and red lines, respectively)
and SF$_{1,2}$ (solid blue and red lines, respectively) are
represented. The thin horizontal line represents the difference in
kinetic energy between the electron and the hole $E_{kin}^{(1,0)}$.
We have used for the Zeeman term an unphysically large value
$g\mu_BB=(1/10)(e^2/\varepsilon \lb)$, for illustration reasons. (b)
Schematic representation of the excitonic modes studied. Notice that
ME$_1$ and ME$_2$ are degenerate in the $N_c\rightarrow \infty$
limit, as well as the SF$_1$ and SF$_2$.}
  \label{NF0}
\end{figure}

\subsection{Modes at filling $\nu=0$}~

\begin{figure}[t]
  \centering
  \includegraphics[width=0.35\textwidth]{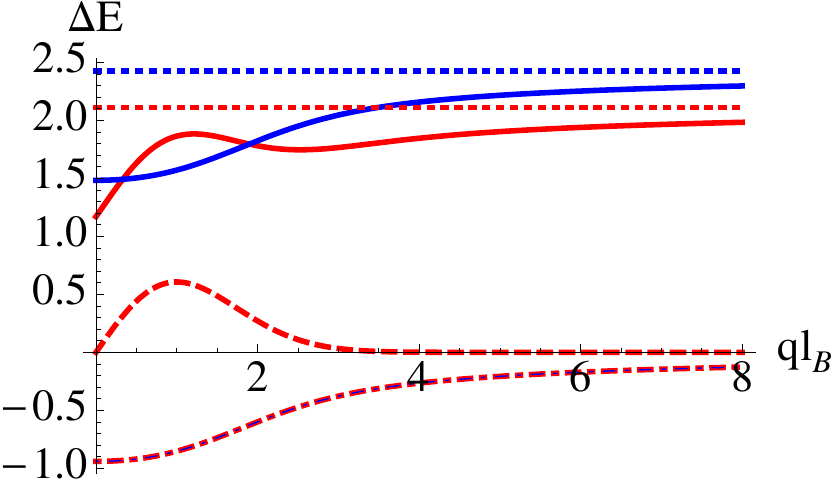}
\caption{(Color online) Decomposition of the ME (full red line) and SF (full blue line) mode for $\nu=0$ into the interaction-related
components, exchange self-energy (dotted line, red for ME and blue for SF), vertex correction (dashed-dotted line) and RPA term
(red dashed line), in units of $e^2/\varepsilon l_B$. The kinetic energy, which yields the same constant offset for both modes, is not taken into account in this
decomposition.}
  \label{NF0decomp}
\end{figure}

At the charge neutrality point (for a filling factor $\nu=0$), the
Fermi level is in the $n=0$ LL (i.e. $N_F=0$), with the
spin-$\downarrow$ branch completely filled ($\nu_{\downarrow}^0=1$)
and an empty spin-$\uparrow$  ($\nu_{\uparrow}^0=0$). The dispersion
of the excitonic modes for this situation is shown in Fig.
\ref{DisperNF0}.  The transitions corresponding to the different
excitations are schematized in Fig. \ref{TransNF0}. To more easily
distinguish between the different modes, we use the notation $\Delta
E_{N_F;\nu^{N_F}_{\downarrow},\nu^{N_F}_{\uparrow}}(q)$. Therefore,
the dispersion of the magnetoexciton modes ME$_{1,2}$, Eq.
(\ref{Eq:ME}) will correspond to the kinetic particle-hole energy
difference plus a renormalization due to electron-electron
interactions, $\Delta E^{ME_{1,2}}_{0;1,0}(q)$, which reads $\Delta
E^{(1,-1/2;0,-1/2)}(q)=\Sigma^{ME_1}_{0;1,0}+V^d_{1,0;1,0}(q)+4V^x_{1,0;1,0}(q)$
for ME$_1$ and $\Delta
E^{(0,+1/2;1,+1/2)}(q)=\Sigma^{ME_2}_{0;1,0}+V^d_{0,-1;0,-1}(q)+4V^x_{0,-1;0,-1}(q)$
for ME$_2$, where the expressions for $\Sigma^{ME_{1,2}}_{0;1,0}$ are
given in Appendix \ref{App:Self}. Here $V^x(q)$ are matrix elements
of the Hartree term, in which a particle-hole pair recombines,
exciting a new particle-hole pair (the usual bubble diagrams). On
the other hand, $V^d(q)$ is the Fock term, which accounts for the
direct interaction of the excited electron and hole (ladder
diagrams). Notice that, due to particle-hole symmetry at this
filling,  $\Delta E^{ME_1}_{0;1,0}(q)=\Delta E^{ME_2}_{0;1,0}(q)$
and the two modes are degenerate. The first thing one notices is
that the dispersion at $q=0$ is shifted with respect to
$E_{kin}^{(1,0)}$ [horizontal line in Fig. \ref{DisperNF0}]. This is
a consequence of the non-applicability of Kohn's theorem in
graphene, as discussed in Sec. \ref{sec:Kohn}, whereas in the 2DEG
the theorem is satisfied due to a cancelation between the exchange
self-energy and the $q=0$ vertex correction, $E_v(q=0)=-E_{exch}$.
Whereas the behavior of the dispersion at short wavelength is
dominated by the exchange self-energy and vertex correction terms
(see Fig. \ref{NF0decomp}), the peak in the dispersion in the
long-wavelength regime is due to the exchange interaction (the RPA
term). Furthermore, it is worth  pointing out that this contribution
rapidly increases as one fills more LLs, as we will see
below. This is a direct consequence of the relativistic quantization
of the graphene LL spectrum, leading to an important LL mixing 
at higher fillings and, as a consequence, building an
unusual particle-hole excitation spectrum.\cite{RFG09}

This RPA contribution is absent, however, in the SF and SW modes
(see Fig. \ref{NF0decomp}). As a consequence, the LL mixing for
these modes is less important and makes the KH
approximation, as the one applied here, a justified method
(especially at strong magnetic fields and for the chemical potential
at or near the zero energy LL). The results for these modes are also
shown in Fig. \ref{DisperNF0}. Electron-electron interactions enter
in the dispersion of the former through the term $\Delta
E^{SF_{1,2}}_{0;1,0}(q)$, which again due to particle-hole symmetry
leads to degenerate modes with contributions $\Delta
E^{(1,+1/2;0,-1/2)}(q)=\Sigma^{SF_1}_{0;1,0}+V^d_{1,0;1,0}(q)$ and
$\Delta
E^{(-1,-1/2;0,+1/2)}(q)=\Sigma^{SF_2}_{0;1,0}+V^d_{0,-1;0,-1}(q)$
respectively. As in the 2DEG, the $q\rightarrow 0$ limit of the
dispersion of these modes is renormalized from the noninteracting
value, $E_{kin}^{(n_e,n_h)}+g\mu_BBS_z$. This makes possible the
study of correlation effects by optical measurements.

On the other hand, Larmor's theorem still applies in graphene, as one may see from the dispersion of the SW mode. This mode has a $q=0$ dispersion equal to the Zeeman splitting $g\mu_BBS_z$, and a contribution due to electron-electron interaction $\Delta E^{(0,+1/2;0,-1/2)}(q)=\Sigma^{SW}_{0;1,0}+V^d_{0,0;0,0}(q)$, which is finite only at non-zero wave-vectors. This implies that, as in the 2DEG, the $g$-factor is not influenced by the Coulomb interaction. The SW disappears if we fill the next LL, for  $\nu=2$, with $N_F=0$, $\nu_{\downarrow}^0=1$, and $\nu_{\uparrow}^0=1$. The dispersions of the ME and SF modes (not shown here) are similar to the previous case with the difference that the degeneracy of the latter is lifted, but only by a constant term equal to the double of the Zeeman energy, in agreement with the arguments of Sec. \ref{sec:Larmor}.

\subsection{Modes at filling $\nu=4$}~

\begin{figure}[t]
  \centering
    \subfigure[]{\label{DisperNF1}\includegraphics[width=0.35\textwidth]{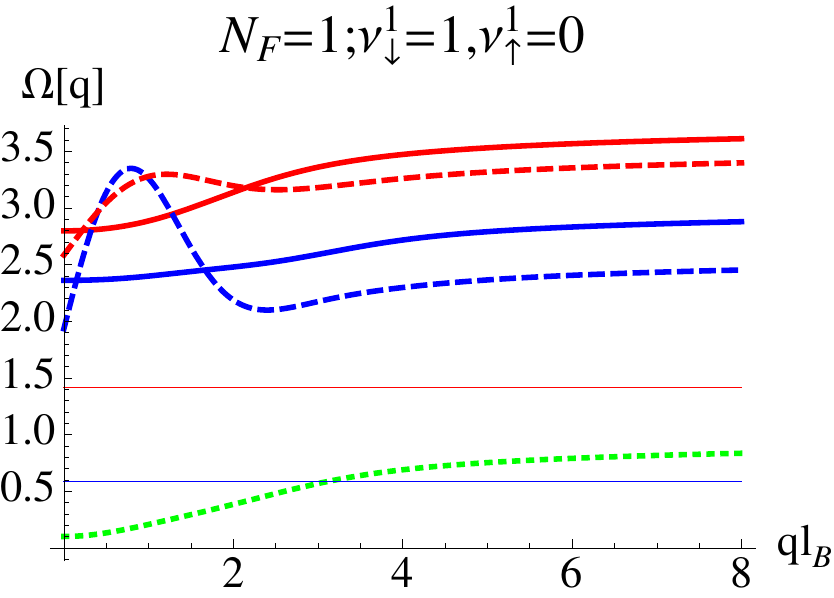}}
     \subfigure[]{\label{TransNF1}\includegraphics[width=0.35\textwidth]{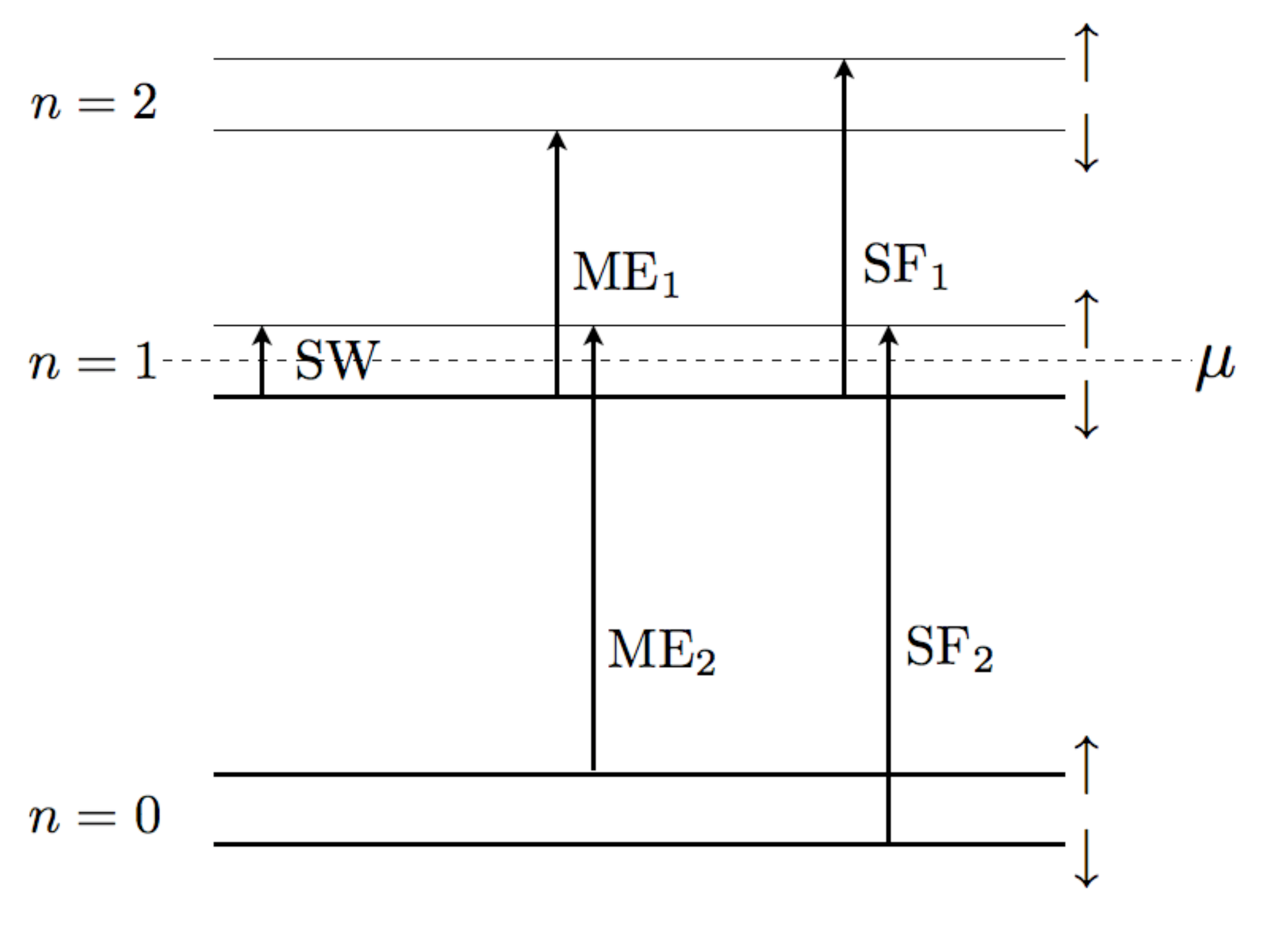}}
      \subfigure[]{\label{SFNF1}\includegraphics[width=0.35\textwidth]{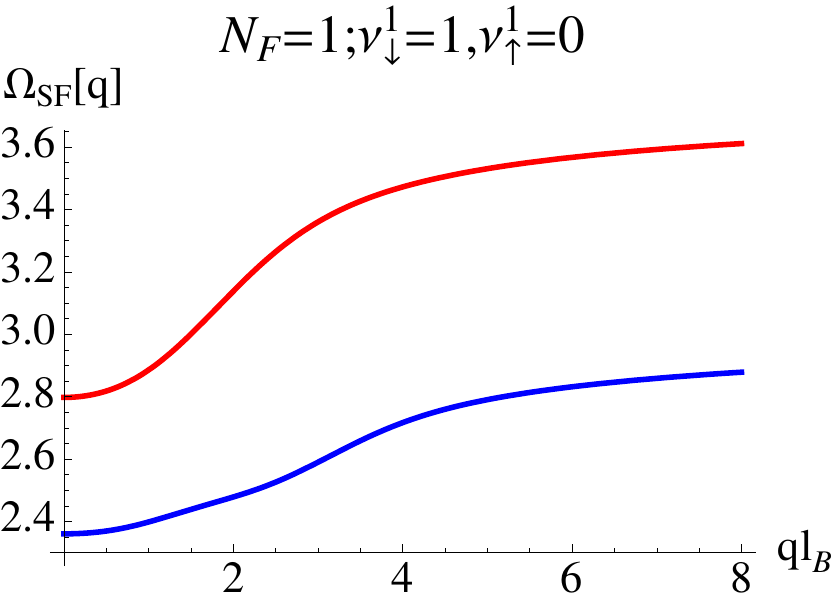}}
 \caption{(a) Same as Fig. \ref{NF0} but for $\nu=4$, with
$N_F=1$, $\nu_{\downarrow}^1=1$ and $\nu_{\uparrow}^1=0$. ME$_{1,2}$ (dashed blue and red lines, respectively) and SF$_{1,2}$ (solid blue and red lines, respectively) are represented. The thin horizontal lines represent the difference in kinetic energy between the electron and the hole  $E_{kin}^{(1,0)}$ and $E_{kin}^{(2,1)}$. (b) Schematic representation of the excitonic modes studied. The degeneracy that occurs at $N_F=0$ is completely lifted at this filling for both, ME and SF modes. For clarity, we show in (c) the SF modes separately.}
  \label{NF1}
\end{figure}

The relativistic nature of the LLs in graphene is clearly visible if
we go beyond $N_F=0$, as shown in Fig. \ref{TransNF1} for a filling
factor of $\nu=4$, with $N_F=1$, $\nu^1_{\downarrow}=1$, and
$\nu^1_{\uparrow}=0$. At this filling, the non-equidistancy of the
LLs lifts the degeneracy of the two ME modes, as well as the two SF
modes. In addition, the exchange contribution to the ME modes, which
leads to the peak in their dispersion, increases as we decrease the
separation between the LLs of the electron and the hole. 
This yields a strong mixing among the different
branches of MEs, as may be seen in  Fig. \ref{DisperNF1}. In fact,
the height of the peak associated with the ME$_1$ [with $n_e=2$ and
$n_h=1$, as represented in Fig. \ref{TransNF1}], is larger than that
of ME$_2$ (with $n_e=1$ and $n_h=0$). This is due to the linear
dispersion of the spectrum, which enhances the quantum effects as we
go to higher filling factors. 
Taking into account that we are showing here only two of the
spin-conserving excitations possible for this filling (the ones
involving the more adjacent LLs to the chemical potential), one can
conclude that no {\it single} MEs will be accessible experimentally
at finite wave-vectors, but a superposition of them. Therefore, the
TDHF method in the strong-field approximation is not valid for the
spin-conserving modes, and the inclusion of a much higher number of
modes is necessary to obtain a reliable result. In fact, this
overlap of different MEs leads to a new set of collective modes, the
linear magneto-plasmons, which have been studied
elsewhere.\cite{RFG09}

Notice that the above arguments are valid only at non-zero values of the wave-vector, whereas the LL mixing effects
are less pronounced at $q=0$, which is the relevant ME energy in magneto-optical experiments.\cite{SH06,DG07,JS07,HS10} However,
as we have mentioned above, also at $q=0$ the ME energy, which is the inter-LL transition energy measured in spectroscopy, is
renormalized due to electron-electron interactions.

The mixing between different contributions is less dramatic for the
SF modes, as one sees in Fig. \ref{SFNF1}, where we show a plot with
only the SF modes at this filling. One notices how the two modes are
clearly decoupled, making the use of the KH approximation more
justified,  because of the absence of the RPA term which is
responsible, in the ME case, for the LL mixing at non-zero values of
the wave-vector. Although not too clearly, it is appreciable that
the number of relative extrema (maxima and minima) in the dispersion
of SF$_1$ (blue line) is higher than for for SF$_2$ (red line). This
is directly related to the node structure of the (hole-) LL
wave-function.\cite{RGF10} These maxima and minima lead to hot spots
in the dispersion that may be detected by Raman spectroscopy
techniques.\cite{EW99}

\section{Renormalization of the chemical potential}\label{Sec:MuEx}

To gain further insight into the effect of electronic interactions in a
graphene flake, we calculate in this section the exchange correction
to the chemical potential,  from a density-matrix approach. This is
the first step toward including electron-electron interaction in the
system. The correction is intrinsically related to the antisymmetry
of the electronic wave-function, which implies, even in the absence
of interactions, a certain amount of correlation between the
positions of two particles with the same spin. Furthermore, its sign
is always negative, due to the fact that it is the interaction of
each electron with the positive charge of its exchange hole. 
One of the effects of Coulomb interaction is a renormalization of
the chemical potential $\mu$, which at zero temperature is the
partial derivative of the total energy with respect to the number of
particles. It contains a contribution from the kinetic energy and also
from interactions. The latter can be written as a mean-field
contribution plus correlation: $\mu=K+\mu^{ex}+\mu^c$, where $K$ is
the kinetic energy, and $\mu^{ex}$ and $\mu^c$ are the exchange and
correlation corrections to the chemical potential, respectively. As usual, the
direct (Hartree) mean field contribution does not appear as it is
compensated by the positively charged background (or neutralizing
background), see the jellium model. Furthermore, the exchange
interaction can lead to a ferromagnetic instability in a dilute
electron gas.\cite{GV05} In graphene, ferromagnetism due to the
exchange interaction between Dirac fermions has also been
studied.\cite{PGC05}  In a magnetic field, $\mu^{ex}$ can be
obtained from the pair correlation function $g(r)$ (see Appendix
\ref{App:CorrFunc} for details of the calculation) as
\begin{equation}\label{Eq:muex}
\mu^{ex}=\bar{n}\int d^2\br\frac{e^2}{\varepsilon r}[g(r)-1]
\end{equation}
where $\bar{n}=4(1+N_c+N_F)/(2\pi\lb^2)$ is the electron density for
graphene in a magnetic field. $N_F$ is the index of the last
occupied LL, related to the filling factor by $\nu=4N_F+2$, and
$N_c$ is a cutoff chosen such that $(4N_c+2)N_B=2N_{u.c.}$, where
$N_B={\cal A}/2\pi l_B^2$ is the degeneracy of each LL, ${\cal A}$
is the surface of the sample, $N_{u.c.}$ is the number of occupied
unit cells in the system, the factor 2 is due to spin
degeneracy, and $4N_c+2$ is the number of filled sub-levels of the
valence band for undoped graphene. $N_c$ is the index of the
last LL in the band (a kind of bandwidth) and is roughly given by
$N_c \approx N_{u.c.}/(2N_B)=2\pi l_B^2/(3\sqrt
3a^2)\approx40000/B[T]$ which is always much greater than 1 in
practise. The fact that $N_c\gg 1$ is just the statement
that, with available magnetic fields, the flux per unit cell is
always much smaller than the flux quantum. In this respect, we are
always in the weak field limit. An exact solution of Eq.
(\ref{Eq:muex}) is possible in the limit $N_c,N_F\gg 1$, as shown in
Eq. (\ref{Eq:muexDM}). This correction would eventually involve a renormalization of $N_F$, this is, a shift 
of the chemical potential as compared to the non-interacting case.

Notice that, contrary to the strong magnetic field assumption done
in the previous section, this is the opposite case, namely the weak
magnetic field limit.  The strong field limit is actually the
KH approximation of weak LL mixing.\cite{KH84} As stated in Sec.
\ref{Sec:KHappr}, the criterion for a weak LL mixing is
$e^2/(\varepsilon l_B \omega_c) \ll 1$. In graphene, because the
fine structure constant $\alpha_G=e^2/\varepsilon v_F$ is of order
one, it means that $k_Fl_B \ll 1$, which means $N_F \approx 0$ (i.e.
$N_F \ll 1$) or in other words $B \gg 20$T. This is the assumption
made in the previous section, whereas in this section we assume the
opposite limit ($B\ll 20$T or $N_F\gg1$). In a standard 2DEG with a
parabolic band, the weak magnetic field limit implies that the
typical Coulomb energy exceeds the cyclotron frequency $\omega_C$.
This allows us to start from the Landau Fermi liquid theory at zero
magnetic field.\cite{AG95} In the case of graphene, this limit is
even more relevant due to the relativistic quantization of the
spectrum into non-equidistant LLs, the relative separation of which
decreases as the energy increases. Therefore, even in a strong
magnetic field, the strength of the Coulomb interaction can be much
higher than the separation between the LLs adjacent to the chemical
potential (the effective cyclotron frequency in graphene) if the
system is sufficiently doped. Further simplification is possible if
$1\ll N_F \ll N_c$. In this limit we obtain (see Appendix
\ref{App:CorrFunc}) that the exchange correction to the Fermi
energy behaves asymptotically as
\begin{equation}\label{Eq:muexDMAsimp}
\mu^{ex}\simeq -\rs \frac{16\sqrt{2}}{3\pi}\sqrt{N_c}.
\end{equation}
This contribution is expected since the energy calculated
above includes the interaction energy of the vacuum of negative
energy particles. It is interesting to compare this leading behavior
of the exchange energy, valid for high filling factors, to the
exchange self-energy obtained in Appendix \ref{App:Self} valid at
low fillings [see e.g. Eqs. (\ref{Eq:SelfNF0})-(\ref{Eq:SelfNF1})
for $N_F=0$ and 1 respectively]. In the two cases we obtain the same
$\sim N_c^{1/2}$ leading behavior.\footnote{Notice that this is also
the behavior found by Aleiner and Glazman for a 2DEG at high filling
factors if we replace $N_c$ by $N_F$. See e. g. Eq. (B28) of Ref.
\onlinecite{AG95} where they consider a single parabolic band (no
need of ultraviolet cutoff).} Furthermore, our results agree
with the exchange contribution calculated for graphene at zero
magnetic field, where a $\Sigma^{ex}\sim-e^2k_c/\varepsilon$
contribution was found, $k_c\sim 1/a$ being an UV cutoff in
momenta.\cite{HHS07} Taking into account  that $N_c\sim (l_B/a)^2$,
our results for graphene in a magnetic field qualitatively agree
with those at $B=0$. Notice that whereas $\mu^{ex}$ diverges as
$N_c^{1/2}$ for the single particle dispersion, the dispersion of a
particle-hole pair diverges only logarithmically, because the terms
proportional to the square root of $N_c$ for each particle cancel
each other, leading to a behavior $E_{exch}\propto \log{N_c}$. This
divergence can be reabsorbed into a renormalization of the Fermi
velocity,\cite{GGV94} and its effect for cyclotron resonance
measurements has been studied in detail by Shizuya.\cite{S10} This
renormalization of the chemical potential due to Coulomb interaction
should affect the scanning single-electron transistor measurements
of compressibility in graphene.\cite{MY07}

\section{Summary and conclusions}\label{Sec:Conc}

In conclusion, we have studied the SF, SW and
ME (or magneto-plasmon) modes in graphene in the
integer QHE regime, in the Kallin-Halperin approximation.
The ME dispersion in a 2DEG is not renormalized
in the long-wavelength limit due to Kohn's
theorem for systems with a parabolic band and Galiean invariance. As
a consequence, the correction due to the direct interaction between
the electron and the hole is neutralized by their difference in
exchange self-energy, $E_v(q=0)=-E_{exch}$, leading to a dispersion
that tends to $m\omega_C$ at zero wave-vector.\cite{PW92} In
graphene, Kohn's theorem does not apply and the dispersion of the ME
modes is renormalized due to many-body effects even at $\bq=0$.

On the other hand, virtual transitions from the vacuum (valence
band) enhance the depolarization term of the spin-conserving ME
dispersion, which enters through the RPA contribution and which
leads to an important LL mixing. We have shown that the mixing is
higher as we increase the LL filling and/or decrease the magnetic
field, invalidating the applicability of KH approximation
for $\nu\ge 2$, which needs to be restricted to the large-field
$N_F=0$ case.\cite{IWFB07,BM08} One of our main conclusions is that, for ME modes,
methods involving more inter-LL transitions than only one
need to be considered in the calculation of the spin-conserving collective
excitations. This superposition of several inter-LL transitions is at the
origin of the strongly-dispersing linear magneto-plasmons, which have been obtained
within an RPA treatment of the electron-electron interactions.\cite{RFG09,RGF10}

In contrast to the spin-conserving ME modes, the depolarization term
is absent in collective excitations where the particle and hole components have 
opposite spin, and the amount of mixing is less
important. Therefore the KH approximation can still be used
for these modes in undoped or slightly doped graphene in a strong
magnetic field. In a 2DEG, the zero-wave-vector limit of the
KH correction of SF modes has a finite contribution, because
$E_v(q=0)=-(1/2)E_{exch}$ in this case. In graphene, the dispersion
of these modes is also renormalized at zero wave-vector and leads to
a correction that could be detected in inelastic light scattering
experiments, by using the same techniques as for a 2DEG.\cite{PW92,KW01,VW06,EW99} In contrast to Kohn's theorem, we have shown that Larmor's theorem
applies to graphene, so that the $q\rightarrow 0$ limit of the SW
dispersion is equal to the Zeeman splitting and the $g$-factor is
independent of many-body interaction, as in a standard
2DEG.\cite{DKW88} In addition, the $g$-factor is also only
weakly affected by band effects in graphene: the effective
$g$-factor was measured to be close to its bare value of 2, see Ref.
\onlinecite{ZK06}.

Finally, we have calculated the exchange shift of the chemical
potential in the weak-magnetic-field limit. We have found that, as
for strong magnetic fields, the exchange correction to the chemical
potential diverges with the ultraviolet cutoff as $\sim N_c^{1/2}$.
However, when the dispersion of an electron-hole pair is considered,
the correction associated with the difference in exchange
self-energy between the particle and the hole, diverges only
logarithmically. This correction leads to a renormalization of
the Fermi velocity that seems to explain some recent
experimental results.\cite{JS07,DG07,HS10}

\begin{acknowledgments}
We thank M. I. Katsnelson for many useful discussions. This work was
funded by ``Triangle de la Physique'' and the EU-India FP-7
collaboration under MONAMI.
\end{acknowledgments}

\appendix

\section{Poles of the response function in the time-dependent Hartree-Fock and strong field approximations}\label{App:Poles}

Within the TDHF approximation, the dispersion relation of the
excitonic modes is defined by the poles of the response function,
which are solutions to the eigenvector equation [see e. g. Ref.
\onlinecite{LK93} for the 2DEG]

\begin{eqnarray}\label{Eq:Poles}
\sum_{\gamma,\delta}\left \{ \delta_{\alpha,\gamma}\delta_{\beta,\delta}[D(\omega)]_{\alpha\beta}^{-1}-\delta_{s^z_{\alpha},s^z_{\gamma}}\delta_{s^z_{\delta},s^z_{\beta}}V^d_{\alpha,\delta;\beta,\gamma}(\bq)\right.&&\nonumber\\
+\left.\delta_{s^z_{\alpha},s^z_{\beta}}\delta_{s^z_{\delta},s^z_{\gamma}}V^x_{\alpha,\delta;\gamma,\beta}(\bq)\right \}B_{\gamma\delta}(\bq)=0,&&
\end{eqnarray}
where $B_{\gamma\delta}(\bq)$ are the basis states, and
$\alpha\equiv (\lambda_{\alpha},n_{\alpha},s^z_{\alpha})$ labels a
particle with band index $\lambda_{\alpha}$, LL $n_{\alpha}$, and
$s^z_{\alpha}$ is the $z$-component of its spin. The sum in Eq.
(\ref{Eq:Poles}) is restricted, in the strong-field approximation
(i.e. weak LL mixing $e^2/(\varepsilon l_B\omega_C)\ll 1$), to pairs
of indices such that $n_{\gamma}-n_{\delta}=n_{\alpha}-n_{\beta}=m$,
and $s^z_{\gamma}-s^z_{\delta}=s^z_{\alpha}-s^z_{\beta}=S^z$.
This is what we call the Kallin-Halperin approximation. The matrix
elements of the two-particle propagator are
\begin{eqnarray}
D_{\alpha,\beta}(\omega)&=&\frac{f_{\alpha}(1-f_{\beta})}{\omega-E_{kin}^{(n_{\beta},n_{\alpha})}-g\mu_BB(s^z_{\beta}-s^z_{\alpha})-E^{exch}_{\beta\alpha}+i\delta}\nonumber\\
&-&\frac{f_{\beta}(1-f_{\alpha})}{\omega-E_{kin}^{(n_{\beta},n_{\alpha})}-g\mu_BB(s^z_{\beta}-s^z_{\alpha})-E^{exch}_{\beta\alpha}-i\delta},\nonumber
\end{eqnarray}
where $f_{\alpha}\equiv\Theta\left(\mu-\lambda_{\alpha}\vf l_B^{-1}\sqrt{2n_{\alpha}}\,\right )$, $\Theta(x)$ being the step function, and $\delta\rightarrow 0^+$. The difference in exchange self-energy between the electron and the hole reads
\begin{eqnarray}
E^{exch}_{\beta\alpha}&=&\Sigma^{\beta}-\Sigma^{\alpha}\nonumber\\
&=&\sum_{\gamma}f_{\gamma}\left[\delta_{s^z_{\beta},s^z_{\gamma}}V^d_{\beta,\beta;\gamma,\gamma}(0)-\delta_{s^z_{\alpha},s^z_{\gamma}}V^d_{\alpha,\alpha;\gamma,\gamma}(0)\right ],\nonumber
\end{eqnarray}
where the direct term is\cite{IWFB07}
\begin{widetext}
\begin{equation}
V^d_{\alpha,\beta;\alpha'\beta'}(\bq)=-\frac{1}{4}(\sqrt{2})^{d_{\alpha,\beta;\alpha',\beta'}}\sum_{\mu,\nu=0}^1b_{\mu}(\lambda_{\alpha})b_{\nu}(\lambda_{\beta})b_{\mu}(\lambda_{\alpha'})b_{\nu}(\lambda_{\beta'})\tilde{u}_{c_{\mu}(\alpha),c_{\nu}(\beta);c_{\mu}(\alpha'),c_{\nu}(\beta')}(\bq),
\end{equation}
\end{widetext}
where $d_{\alpha,\beta;\alpha',\beta'}=\delta_{n_{\alpha},0}+\delta_{n_{\beta},0}+\delta_{n_{\alpha'},0}+\delta_{n_{\beta'},0}$, $b_0(\lambda)=1$, $b_1(\lambda)=\lambda$, $c_0(\alpha)=|n_{\alpha}|$ and $c_1(\alpha)=|n_{\alpha}|-1$, and
\begin{equation}
\tilde{u}_{\alpha,\beta;\alpha'\beta'}(\bq)=\frac{1}{l_B^2}\int d\br\, v(\br-l_B^2\uz\times \bq){\cal F}_{\alpha',\beta'}^*(\br){\cal F}_{\alpha,\beta}(\br),
\end{equation}
$v(\br)=e^2/\varepsilon r$ being the Coulomb potential and
\begin{eqnarray}
{\cal F}_{\alpha,\beta}(\br)&=&\frac{1}{\sqrt{2\pi}}\frac{1}{2^{|m|/2}}\frac{n_<!}{\sqrt{n_{\alpha}!n_{\beta}!}}e^{-im\phi}\sgn(m)^m\nonumber\\
&&\times\left (\frac{r}{l_B}\right )^{|m|}L_{n_<}^{|m|}\left(\frac{r^2}{2l_B^2}\right )e^{-\frac{r^2}{4l_B^2}},
\end{eqnarray}
where $n_<=\min(n_{\alpha},n_{\beta})$, $m=n_{\alpha}-n_{\beta}$ and $e^{i\phi}=(x+iy)/|x+iy|$, and $\sgn (m)^m=1$ for $m=0$. The exchange matrix elements read
\begin{widetext}
\begin{equation}
V^x_{\alpha,\beta;\alpha'\beta'}(\bq)=-\frac{1}{4}(\sqrt{2})^{d_{\alpha,\beta;\alpha',\beta'}}\sum_{\mu,\nu=0}^1b_{\mu}(\lambda_{\alpha})b_{\nu}(\lambda_{\beta})b_{\mu}(\lambda_{\alpha'})b_{\nu}(\lambda_{\beta'})\tilde{v}_{c_{\mu}(\alpha),c_{\nu}(\beta);c_{\mu}(\alpha'),c_{\nu}(\beta')}(\bq),
\end{equation}
\end{widetext}
where
\begin{equation}
\tilde{v}_{\alpha,\beta;\alpha'\beta'}(\bq)=\frac{1}{l_B^2}\frac{2\pi e^2}{\varepsilon q}{\cal F}_{\alpha',\beta'}^*(l_B^2\uz\times\bq){\cal F}_{\alpha,\beta}(l_B^2\uz\times\bq)
\end{equation}

\section{Exchange self-energy contributions}\label{App:Self}

In this appendix we give analytical expressions for the contributions to $\Delta E$ associated to the difference in exchange self-energy between the electron and the hole. For $N_F=0;\nu_{\downarrow}^0=1,\nu_{\uparrow}^0=0$, using the notation $\Sigma_{N_F;\nu_{\downarrow}^{N_F},\nu_{\uparrow}^{N_F}}$, we obtain
\begin{equation}\label{Eq:SelfME1}
\Sigma^{ME_1}_{0;1,0}=\rs\left [\frac{3}{4} \sqrt{\frac{\pi }{2}}+ \sum _{n=1}^{N_c} \frac{\left(4 \sqrt{n}-3\right) \Gamma \left(n-\frac{1}{2}\right)}{16 \sqrt{2} \Gamma (n+1)} \right ]
\end{equation}
for ME$_1$ and
\begin{equation}
\Sigma^{ME_2}_{0;1,0}=\rs \sum _{n=1}^{N_c}\frac{\left(4 \sqrt{n}+3\right) \Gamma \left(n-\frac{1}{2}\right)}{16 \sqrt{2} \Gamma (n+1)}
\end{equation}
for ME$_2$, where $N_c$ is a high-energy cutoff. Notice that $\Sigma^{ME_1}_{0;1,0}=\Sigma^{ME_2}_{0;1,0}$ in the limit $N_c\rightarrow \infty$. The contributions to the spin flip modes are
\begin{eqnarray}\label{Eq:SelfSF1}
\Sigma^{SF_1}_{0;1,0}&=&\Sigma^{ME_1}_{0;1,0}-V_{1,1;0,0}^d(0)\\ \nonumber
&=&\rs\left [\sqrt{\frac{\pi }{2}}+ \sum _{n=1}^{N_c} \frac{\left(4 \sqrt{n}-3\right) \Gamma \left(n-\frac{1}{2}\right)}{16 \sqrt{2} \Gamma (n+1)} \right ]
\end{eqnarray}
for SF$_1$ and
\begin{eqnarray}
\Sigma^{SF_2}_{0;1,0}&=&\Sigma^{ME_2}_{0;1,0}-V_{-1,-1;0,0}^d(0)\\ \nonumber
&=&\rs \left[ \frac{1}{4}\sqrt{\frac{\pi}{2}}\sum _{n=1}^{N_c}\frac{\left(4 \sqrt{n}+3\right) \Gamma \left(n-\frac{1}{2}\right)}{16 \sqrt{2} \Gamma (n+1)} \right ]
\end{eqnarray}
for SF$_2$. Again, $\Sigma^{SF_1}_{0;1,0}=\Sigma^{SF_2}_{0;1,0}$ as the cutoff $N_c$ tends to infinity. On the other hand, the contribution for the SW mode is $\Sigma^{SW}_{0;1,0}=e^2/(\varepsilon l_B) \sqrt{\pi/2}$, which is cutoff independent. The contributions for $N_F=0;\nu_{\downarrow}^0=1,\nu_{\uparrow}^0=1$ can be expressed in terms of the previously given $\Sigma^{ME_{1,2}}_{0;1,0}$, as $\Sigma^{ME_1}_{0;1,1}=\Sigma^{ME_2}_{0;1,1}=\Sigma^{ME_1}_{0;1,0}$ for the ME modes and $\Sigma^{SF_1}_{0;1,1}=\Sigma^{SF_2}_{0;1,1}=\Sigma^{ME_1}_{0;1,0}$ for the SF modes.

Finally, the contributions for $N_F=1;\nu_{\downarrow}^1=1,\nu^1_{\uparrow}=0$, shown in Fig. \ref{NF1}, are
\begin{eqnarray}
\Sigma^{ME_1}_{1;1,0}&=&\rs\left[\frac{1}{128} \left(37 \sqrt{2}-8\right) \sqrt{\pi }\right.\nonumber\\
\nn
&+&\left.\sum _{n=1}^{N_c} \frac{\left[8 \sqrt{n} \left(4 n-2 \sqrt{n}-3\right)+3\right]\Gamma \left(n-\frac{3}{2}\right) }{128 \sqrt{2}\Gamma (n+1)}\right]\\ ~
\end{eqnarray}
for the ME$_1$ mode, whereas $\Sigma^{ME_2}_{1;1,0}=\Sigma^{ME_1}_{0;1,0}$ as given in Eq. (\ref{Eq:SelfME1}). For the spin-flip modes we have
\begin{eqnarray}
\Sigma^{SF_1}_{1;1,0} &=& \rs\left[\frac{3}{4} \sqrt{\frac{\pi }{2}}\right.\\
\nn
&&\left. +\sum _{n=1}^{N_c} \frac{8 \sqrt{n} \left(4 n-2 \sqrt{n}-3\right)+3 }{128 \sqrt{2}}\frac{\Gamma \left(n-\frac{3}{2}\right)}{ \Gamma (n+1)}\right ]
\end{eqnarray}
for SF$_1$ while the contribution associated to the second mode is $\Sigma^{SF_2}_{1;1,0}=\Sigma^{SF_1}_{0;1,0}$ and coincides with Eq. \ref{Eq:SelfSF1}. Finally, the $N_c$-independent contribution to the SW mode is $\Sigma^{SW}_{1;1,0}=\rs\frac{11}{16}\sqpi$.

In the following, we calculate the exchange energy of the system at low fillings. First, one notices that the exchange self-energy for undoped graphene ($N_F=0; \nu^0_{\downarrow}=1,\nu^0_{\uparrow}=0$) [and similarly for the filling ($N_F=-1; \nu_{\downarrow}^{-1}=1, \nu_{\uparrow}^{-1}=1$)]  can be calculated as $\Sigma^{ex}=\sum_{n=-N_c}^{-1}V_{0,0;n,n}^d(0)=-e^2/(2\sqrt{2}\varepsilon l_B) \sum_{n=1}^{N_c}\Gamma\left(n+\frac{1}{2}\right)/\Gamma\left(n+1\right)$, which can be summed up exactly to give
\begin{equation}
\Sigma^{ex}=-\rs\frac{1}{2\sqrt{2}}\left[-\sqrt{\pi} +2\frac{\Gamma\left(N_c+\frac{3}{2}\right)}{\Gamma(N_c+1)}\right]
\end{equation}
For $N_c\gg 1$ we obtain the asymptotic behavior
\begin{equation}\label{Eq:SelfNF0}
\Sigma^{ex}\simeq -\rs \frac{1}{2\sqrt{2}}\left [-\sqrt{\pi}+2\sqrt{N_c}+O\left(\frac{1}{N_c} \right )^{1/2} \right ]
\end{equation}
It is useful to express this result by substituting $\sqrt{N_c}$ by its magnetic-field dependence $\sqrt{N_c}\sim l_B/a$. By doing so, we obtain
\begin{equation}\label{Eq:SelfNF0b}
\Sigma^{ex}=-\frac{e^2}{\varepsilon
a}\left[\#+\#\frac{a}{l_B}+O\left(\frac{a}{l_B}\right)^2\right],
\end{equation}
where $\#$ stands for some numerical prefactor and $a/l_B=0.006\sqrt{B[T]}$. We clearly see that the dominant term, as in Eq. (\ref{Eq:muexDMAsimp}), is magnetic-field independent. 

A similar result is obtained for doped graphene up to the first LL of the conduction band. If ($N_F=1; \nu_{\downarrow}^1=1,\nu^1_{\uparrow}=0$) or ($N_F=0; \nu_{\downarrow}^0=1,\nu_{\uparrow}^0=1$), then $\Sigma^{ex}$ is computed as
\begin{eqnarray}
\Sigma^{ex}&=&\sum_{n=-N_c}^0V_{1,1;n,n}^d(0)\nonumber\\
&=&V_{1,1;n,n}^d(0)+\rs\sum_{n=1}^{N_c}\frac{1+4\sqrt{n}-8n}{16\sqrt{2}}\frac{\Gamma \left(n-\frac{1}{2}\right)}{\Gamma (n+1)}\nonumber
\end{eqnarray}
Taking $N_c\gg 1$ we obtain the limiting result
\begin{eqnarray}\label{Eq:SelfNF1}
\Sigma^{ex}\simeq -\rs&\times& \left [ \frac{1}{8}\sqrt{\frac{\pi}{2}}+\frac{1}{\sqrt{2}}\sqrt{N_c}\right.\nonumber\\
& -& \left.\frac{4}{16\sqrt{2}}\left (1.0646 +\gamma +\ln N_c \right ) \right ],
\end{eqnarray}
where $\gamma$ is the Euler constant and we have approximated $\sum_{n=1}^{N_c}\left( \sqrt{n}\Gamma(n-1/2)/n!-n^{-1}\right )\simeq 1.0646$. Eq. (\ref{Eq:SelfNF1}) could be accordingly expressed in the form of Eq. (\ref{Eq:SelfNF0b}), and the result would be the similar as before: the leading term in the exchange contribution to the chemical potential does not depend on the magnetic field.

\section{Correlation function}\label{App:CorrFunc}

The one-particle density matrix for the $K$ valley (labeled here by $+$) can be defined as
$
\rho_+(\br,\brp)=\sum_{\sigma}\sum_{\lambda,n}\rho_{+,\la,n}(\br,\brp)
$
in terms of the density matrix of the $n$-th LL of the $\la$ band of the $K$ valley
$
\rho_{+,\la,n}(\br,\brp)=\sum_k\Psi_{\lambda
nk}^{+\dagger}(\br)\Psi_{\lambda nk}^+(\brp),
$
where $\Psi_{\lambda nk}^+(\br)$ are the $K$-valley LL wave-function. The wave-function for graphene in a magnetic field can be constructed from the corresponding nonrelativistic LL wave-functions of a 2DEG with a parabolic band dispersion. In the Landau gauge, where the vector potential is
$\vec{\mathbf{A}}=(0,Bx,0)$, they can be written as
\begin{equation}\label{Eq:WF-LG}
\Psi_{nk}^+(\br)=\frac{1}{\sqrt{L}}e^{-iky}   \left(
                                                     \begin{array}{c}
                                                       -i\lambda 1_n^*\phi_{n-1,k}(x) \\
                                                       2_n^*\phi_{n,k}(x) \\
                                                       0 \\
                                                       0 \\
                                                     \end{array}
                                                   \right)
\end{equation}
for the $K$ ($+$) valley, and
\begin{equation}
\Psi_{nk}^-(\br)=\frac{1}{\sqrt{L}}e^{-iky}   \left(
                                                     \begin{array}{c}
                                                       0 \\
                                                        0 \\
                                                       2_n^*\phi_{n,k}(x) \\
                                                       -i\lambda 1_n^*\phi_{n-1,k}(x) \\
                                                     \end{array}
                                                   \right)
\end{equation}
for the $K^{\prime}$ ($-$) valley, where
\begin{equation}
\phi_{n,k}(x)=\frac{1}{\sqrt{2^nn!\sqrt{\pi}l_B}}e^{-z^2/2}H_n(z).
\end{equation}
In the previous expression $z=x-kl_B^2/l_B$ and $H_n$ are Hermite polynomial, and we have defined $1_n^*=\sqrt{(1-\delta_{n,0})/2}$ and $2_n^*=\sqrt{(1+\delta_{n,0})/2}$. One obtains therefore
\begin{eqnarray}
\rho_{+,\la,n}(\br,\brp)\!&\!=\!&\!\frac{1}{2\pi l_B^2}e^{-\frac{i(y-y')(x+x')}{2l_B^2}}e^{-\frac{|\br-\brp|^2}{4l_B^2}}\nonumber\\
&\!\!\!\times\!\!\!&\!\!\!\left[ 1_n^{*2}L_{n-1}\!\!\left(\frac{|\br-\brp|^2}{2l_B^2}\right
)+2_n^{*2}L_{n}\!\!\left(\frac{|\br-\brp|^2}{2l_B^2}\right )\right ].\nonumber
\end{eqnarray}
The sum $\sum_{\la,n}\rho_{+,\la,n}(\br,\brp)$ in the band and LL indices is decomposed into an inter- and an intra-band contributions
\begin{equation}
\sum_{n=1}^{N_c}\rho_{+,\la=-1,n}(\br,\brp)+\sum_{n=1}^{N_F}\rho_{+,\la=+1,n}(\br,\brp).
\end{equation}
Furthermore, it can be checked that $\sum_{n=1}^{N_0}1^{*2}_nL_{n-1}^0(x)=(1/2)L_{N_0-1}^1(x)$, and $\sum_{n=1}^{N_0}2^{*2}_nL_{n}^0(x)=(1/2)L_{N_0}^1(x)$, where $N_0=N_c,N_F$. Therefore, neglecting the Zeeman splitting, we obtain for the $K$-valley one-particle density-matrix:
\begin{widetext}
\begin{equation}
\rho_+(\br,\brp)=\frac{1}{2\pi\lb^2}e^{-\frac{i(y-y')(x+x')}{2l_B^2}}e^{-\frac{|\br-\brp|^2}{4l_B^2}}\left[ L_{N_c-1}^1\left(\frac{|\br-\brp|^2}{2l_B^2}\right
)+L_{N_c}^1\left(\frac{|\br-\brp|^2}{2l_B^2}\right)+(N_c\rightarrow N_F)\right ],
\end{equation}
\end{widetext}
where $(N_c\rightarrow N_F)$ indicates the replacement of $N_c$ by $N_F$. Considering the $K'$-valley contribution, the one-particle density-matrix is obtained as $\rho(\br,\brp)=2\rho_+(\br,\brp)$. From this, one can obtain the pair correlation function $g(\br,\brp)$, which is defined as the
normalized probability of finding an electron at position $\br$
given that, at the same time, there is another electron at position
$\brp$. It can be expressed in terms of the density matrix
as\cite{GV05}
$
g(\br,\brp)=1-|\rho(\br,\brp)|^2/[n(\br)n(\brp)]
$
where $n(\br)\equiv\rho(\br,\br)=4(1+N_c+N_F)/(2\pi\lb^2)$ is the electron density and we have used the fact that $L_n^{\alpha}(0)=(n+\alpha)!/(n!\alpha!)$. Setting $\brp=0$ we find
\begin{widetext}
\begin{equation}\label{Eq:PairCorr}
g(r)=1-\frac{1}{N^2}\left \{2e^{-\frac{r^2}{4l_B^2}}\left [1+L_{N_c-1}^1\left(\frac{r^2}{2l_B^2}\right)+L_{N_c}^1\left(\frac{r^2}{2l_B^2}\right) +(N_c\rightarrow N_F)\right ]\right \}^2,
\end{equation}
\end{widetext}
where $N\equiv 2\pi\lb^2n(\br)$. By using the asymptotic expression $e^{-x/2}L_{n-1}^1(x)\simeq \sqrt{n/x}J_1\left(2\sqrt{xn}\right)$, where $J_1(x)$ is a Bessel function of the first kind, valid for $n\gg 1$, we obtain for $N_F,N_c\gg 1$:
\begin{equation}\label{Eq:PairCorrAppr}
g(r)\simeq 1-\frac{4}{N^2}\left [ e^{-\frac{r^2}{4l_B^2}}+\psi(N_c,r)+\psi(N_F,r)\right]^2
\end{equation}
where $\psi(n,r)=2nJ_1\left(r\lb^{-1}\sqrt{2n}\right )/r\lb^{-1}\sqrt{n/2}$ and we have approximated $N_c-1\simeq N_c$ and $N_F-1\simeq N_F$. Using Eq. (\ref{Eq:PairCorrAppr}) into Eq. (\ref{Eq:muex}), with $ n(\br)\equiv\bar{n}$ being the electron density in the isotropic case, we can obtain an expression for the exchange energy per particle in the large $N_c,N_F$ limit with the exact solution
\begin{widetext}
\begin{eqnarray}\label{Eq:muexDM}
\mu^{ex}&=&-\frac{e^2}{\varepsilon \lb}\frac{4}{3\pi}\frac{1}{N_c}\left\{4 \sqrt{2} \sqrt{N_c}\left[\left(N_F-N_c\right) K\left(\frac{N_F}{N_c}\right)+\left(N_c+N_F\right) E\left(\frac{N_F}{N_c}\right)\right]+4 \sqrt{2} \left(N_c^{3/2}+N_F^{3/2}\right)\right.\nonumber\\
&&+\left. 6 \pi ^{3/2} e^{-N_c} N_c\left[I_0\left(N_c\right)+I_1\left(N_c\right)\right] + 6 \pi ^{3/2} e^{-N_F} N_F\left[I_0\left(N_F\right)+I_1\left(N_F\right)\right]+\frac{3}{2\sqrt{2}}\pi^{3/2}    \right \}
\end{eqnarray}
\end{widetext}
where $K(n)$ and $E(n)$ are elliptic integrals of first and second kind, respectively, and $I_n(z)$ are the modified Bessel functions of the first kind.

\bibliography{BibliogrGrafeno}

\end{document}